\documentclass[usenatbib]{mnras}
\usepackage{natbib}
\usepackage{apjfonts}
\usepackage{amsmath}
\usepackage[table]{xcolor}

\usepackage{epsfig}
\usepackage{graphicx}
\usepackage{amssymb}
\usepackage{aas_macros}
\usepackage{color}

\newcommand{\oversim}[2]{\protect{\mbox{\lower0.5ex\vbox{%
   \baselineskip=0pt\lineskip=0.2ex
   \ialign{$\mathsurround=0pt #1\hfil##\hfil$\crcr#2\crcr\sim\crcr}}}}} 
\newcommand{\bb}[1]{\ifmmode \mbox{\boldmath $ #1$} \else  \mbox{\boldmath $#1$} \fi}
\catcode`\"=\active\let"=\"  

\def\3{{\ss} }

\def\c12{{1\over 2}}

\def\d{{\rm d}}   
   
\def\plusplus{\raise 0.3ex\hbox{${\scriptstyle ++}$}{}}

\def\and{{{\rm M}31}}

\bibliography{biblio}

\usepackage{array}
\newcolumntype{L}[1]{>{\raggedright\let\newline\\\arraybackslash\hspace{0pt}}m{#1}}
\newcolumntype{C}[1]{>{\centering\let\newline\\\arraybackslash\hspace{0pt}}m{#1}}
\newcolumntype{R}[1]{>{\raggedleft\let\newline\\\arraybackslash\hspace{0pt}}m{#1}}

\begin{document}   
\title[Orbital scattering in clumpy potentials]{Orbital scattering by random interactions with extended substructures}

\author[Jorge Pe\~{n}arrubia]{Jorge Pe\~{n}arrubia$^{1,2}$\thanks{jorpega@roe.ac.uk}\\
  $^1$Institute for Astronomy, University of Edinburgh, Royal Observatory, Blackford Hill, Edinburgh EH9 3HJ, UK\\
  $^2$Centre for Statistics, University of Edinburgh, School of Mathematics, Edinburgh EH9 3FD, UK
}
\maketitle  

\begin{abstract}
  This paper presents $N$-body and stochastic models that describe the motion of tracer particles in a potential that contains a large population of extended substructures.
  Fluctuations of the gravitational field induce a random walk of orbital velocities that is fully specified by drift and diffusion coefficients. 
  In the impulse and local approximations the coefficients are computed analytically from the number density, mass, size and relative velocity of substructures without arbitrary cuts in forces or impact parameters.
  The resulting Coulomb logarithm attains a well-defined geometrical meaning, $\ln(\Lambda)=\ln (D/c)$, where $D/c$ is the ratio between the average separation and the individual size of substructures.
 Direct-force and Monte-Carlo $N$-body experiments show excellent agreement with the theory if substructures are sufficiently extended ($c/D\gtrsim 10^{-3}$) and not spatially overlapping ($c/D\lesssim 10^{-1}$).
However, close encounters with point-like objects ($c/D\ll 10^{-3}$) induce a heavy-tailed, non-Gaussian distribution of high-energy impulses that cannot be described with Brownian statistics.
In the point-mass limit ($c/D\approx 0$) the median Coulomb logarithm measured from $N$-body models deviates from the theoretical relation, converging towards a maximum value $\langle \ln(\Lambda)\rangle \approx 8.2$ independently of the mass and relative velocity of nearby substructures.
\end{abstract}   


\section{Introduction}\label{sec:intro}
Describing the motion of tracer particles in a clumpy potential is a key problem in gravitational dynamics with a wide range of applications in Astronomy and plasma physics.  Unfortunately, it typically involves a very large number of coupled equations of motion, which makes it a difficult mathematical problem to tackle. In the last decades, considerable effort has been directed towards devising computational $N$-body methods that provide approximate, albeit tractable, solutions (e.g. Heggie \& Hut 2003). Yet, even with the aid of numerical methods some astronomical problems lie beyond current computational capabilities. For example, cold dark matter models predict a population of $\sim 10^{15}$ dark matter clumps in the halo of Milky Way-like galaxies with masses that span $\sim 18$ orders of magnitude (e.g. Diemand et al. 2005; Springel et al. 2008). To date, following self-consistently the formation and dynamical evolution of these objects represents a forbidding computational challenge (van den Bosch 2017, van den Bosch \& Ogiya 2018; Errani \& Pe\~narrubia 2019).

Given their very large number of degrees of freedom, gravitational systems are particularly well suited for statistical methods. Here, one must abandon the Newtonian's approach of solving the phase-space trajectories of individual particles from deterministic equations of motion, and focus instead on a statistical description of the response of macroscopic objects to repeated interactions with a very large number of low-mass particles. It was Chandrasekhar (1941b) who led the first attempt to construct a statistical theory of stellar encounters using a method originally devised by Holtsmark (1919) to explain the random motion of charged particles in a plasma. In his theory, Chandrasekhar computes the probability $p({\mathbfit F})$ that a tracer star moving in an homogeneous sea of point-masses experiences a combined force ${\mathbfit F}$ within the interval ${\mathbfit F},{\mathbfit F}+\d{\mathbfit F}$, and derives the average squared velocity impulse $\langle |\Delta {\mathbfit v}|^2\rangle$ that would result from a large number of force fluctuations within a short time interval $t$. Due to the singular force induced by point-masses at arbitrarily-close distances, the theory yields a velocity variation that diverges logarithmically as $\langle |{\Delta \mathbfit v}|^2\rangle \propto \ln (\Lambda)$, where $\Lambda=D/r_{\rm min}$ is the so-called Coulomb logarithm, $D$ is the average intra-particle distance, and $r_{\rm min}$ is some arbitrary truncation in the spatial distribution of nearby particles. Unfortunately, the Coulomb logarithm is an ill-defined quantity that reflects a real underlying issue in the stochastic analysis of orbital scattering.

Chandrasekhar's treatment of stellar collisions assumes that a test star has no influence on the background medium. In essence, the dynamic sea of point-masses plays the role of thermal reservoir in statistical mechanics, where the force fluctuations induced by the bath are independent of the motion of the test particle. The case where the particle {\it does} influence the surrounding medium is considerably more convoluted, to the point that it remains an outstanding problem in stellar dynamics after decades of research.
As a massive object moves through an homogeneous distribution of low-mass particles, it experiences two drag forces simultaneously: the first one is a diffusive component that arises from two-body encounters with neighbour stars. By symmetry, the net force points in the direction opposite to the particle motion (Chandrasekhar 1943). The second is due to the self-gravity of the massive particle, which causes an additional friction by polarization typically known as a `gravitational wake' (Weinberg 1989; Nelson \& Tremaine 1999). Both forces are proportional to the Coulomb logarithm that results from averaging over individual encounters, $\ln(\Lambda)=\ln (b_{\rm max}/b_{\rm min})$, which diverges at large and small impact parameters (Chandrasekhar 1943). The minimum impact parameter $b_{\rm min}$ is typically set equal to the distance where the deflection angle is $90^\circ$. Alas, the appropriate choice of maximum impact parameter is controversial. Chandrasekhar (1943) and Kandrup (1980) advocate terminating the integration over impact parameters at the typical interstellar separation, whereas Cohen, Spitzer \& Routly (1950) argue that the integration should include all impact parameters up to the characteristic size of the stellar system (or the Debye length in a plasma). The latter case would cast doubt on two central tenets of Chandrasekhar (1943) theory, namely that dynamical friction can be treated locally, and that interactions can be approximated as a Markovian sequence of two-body encounters of negligible duration.
Numerical studies of satellite decay have shown that Chandrasekhar's equations are accurate enough if the Coulomb logarithm remains as a free parameter to be fitted to $N$-body data (e.g. van den Bosch et al. 1999; Colpi, Mayer \& Governato 1999; Pe\~narrubia et al. 2004; Karl et al. 2015). Unfortunately, the best-fitting values depend on the code parameters and the number of particles in the simulation. For instance, Prugniel \& Combes (1992) and Whade \& Donner (1996) find that dynamical friction is artificially increased due to numerical noise if the particle number is small. Given the number of unresolved theoretical and technical difficulties, it is not surprising that a theory for dynamical friction free from divergences remains elusive to this date (e.g. Just \& Pe\~narrubia 2005 and references therein). 

The necessity to introduce arbitrary truncations in the statistical treatment of stellar encounters has recently motivated alternative methods to describe collisional evolution of self-gravitating systems in which relaxation does not occur only through two-particle scattering and is enhanced by self-gravity.
For example, Fouvry \& Bar-Or (2018) apply the so-called $\eta$-formalism, in which the orbital evolution of the test star is described by a Fokker-Planck equation in action-angle space. The analysis shows that fluctuations of the potential (external and self induced) drive the long-term evolution of the integrals of motion. Following a similar goal, Hamilton et al. (2018) use the inhomogeneous Balescu-Lenard (BL) equation to describe the long-term evolution of inhomogeneous self-gravitating objects. The BL equation has no ill-defined parameter and, unlike the classical theory, includes self-gravity.
Both, the $\eta$-formalism and the BL equation yield the same diffusion coefficients in the case where the bath particles can self interact and the system is allowed to react collectively. 
In agreement with Weinberg (1993), 
these authors find that the classical theory underestimates the rate of relaxation through neglect of self-gravity and non-local resonances. Unfortunately, they conclude that the BL equation does not currently provide a viable alternative to classical theory owing to the difficult evaluation of the infinite sum over interactions.

This paper re-visits the original theory of random force fluctuations of Chandrasekhar (1941b), and extends the analysis to a large population of {\it extended substructures} in dynamical equilibrium within the host potential. It is shown that by considering extended objects rather than point-masses the need of an ad-hoc truncation at strong forces vanishes. As a first step, Section~\ref{sec:fluctu} computes the spectrum of force fields generated by an homogeneous distribution of Hernquist (1990) spheres with known mass and size functions using the technique presented in Pe\~narrubia (2008; hereinafter Paper I). 
By construction, the combined force ${\mathbfit F}=\sum_{i=1}^N{\mathbfit f}_i$ acting on the tracer particle is isotropic and follows a static probability distribution, $p({\mathbfit F})$.
Following Pe\~narrubia (2019; hereinafter Paper II), Section~\ref{sec:kicks} adopts a Brownian motion formalism to derive the distribution of velocity impulses associated with the spectrum of force fluctuations $p({\mathbfit F})$, which reduces the problem of orbital scattering to the computation of drift and diffusion coefficients, $\langle \Delta {\mathbfit v}\rangle$ and $\langle |\Delta {\mathbfit v}|^2\rangle$, respectively. Both coefficients can be calculated analytically under two main assumptions (i) {\it local approximation}: the average separation between substructures is much smaller than the scale-length on which the number density varies, $D\ll |\nabla n /n|^{-1}$, such that $n$ is approximately constant at the location of the test particle, and (ii) {\it impulse approximation}: the fluctuations of the combined force (${\mathbfit F}$) have mean life-times ($T_{\rm ch}$) that are much shorter than the dynamical time of the tracer particle about the host potential, $T_{\rm ch}\ll t_{\rm dyn}=r/v$. For substructures with a given mass ($M$) and size ($c$), these simplifications render a theory that only depends on local quantities: the average distance between substructures ($D$), and their relative velocity with respect to the tracer particle ($\sqrt{\langle v^2\rangle}$).
Furthermore, the Coulomb logarithm acquires a well-defined physical meaning, $\ln(\Lambda)=\ln (D/c)$, i.e. the logarithm of the ratio between the average distance between substructures and their individual sizes (i.e. the two relevant scale-lengths of the system).
Section~\ref{sec:nbody} 
presents a number of direct $N-$body and Monte-Carlo experiments that examine the validity of the analytical expressions as a function of the number of objects in the substructure population ($N$) and their size-to-distance ratio ($c/D$).
The divergent behaviour of $\ln(\Lambda)$ in the point-mass limit $c/D\to 0$ is investigated in detail with the aid of direct $N$-body calculations in Section~\ref{sec:disc}. The main findings of the paper are summarized in Section~\ref{sec:sum}, together with a brief description of future applications of the theory.

\section{Stochastic models}\label{sec:sto}
This Section summarizes the main stochastic techniques applied in this paper to construct a Brownian theory for tracer particles surrounded by an homogeneous sea of moving objects. As a first step, Section~\ref{sec:fluctu} follows the steps outlined in Paper I to compute the function $p({\mathbfit F})$, which determines the probability that a tracer particle experiences a combined force ${\mathbfit F}=\sum_{i=1}^N {\mathbfit f}_i$ within the interval ${\mathbfit F},{\mathbfit F}+\d{\mathbfit F}$. Next, following the methodology introduced in Paper II, Section~\ref{sec:kicks} derives the coefficients of the 
distribution of velocity impulses in time intervals that are much shorter (`static limit') or much longer (`dynamic limit') than the average duration of force fluctuations. 

\subsection{Fluctuations of the field}\label{sec:fluctu}
The equations of motion that determine the trajectory of a tracer particle in a clumpy environment can be expressed as (Chandrasekhar 1941a)
\begin{align}\label{eq:eqmot}
\frac{{\d^2 \mathbfit r}_\star}{\d t^2}=-\nabla\Phi_g({\mathbfit r}_\star) + \sum_{i=1}^{N}\, {\mathbfit f}_i({\mathbfit r}_\star),
\end{align}
where $\Phi_g$ is the mean-field (i.e. `smooth') gravitational potential of the host galaxy, and
\begin{align}\label{eq:F}
{\mathbfit F}\equiv \sum_{i=1}^{N}\,{\mathbfit f}_i,
\end{align}
is the specific force induced by a population of extended substructures. The combined force ${\mathbfit F}$ depends on the instantaneous relative position of $N\gg 1$ clumps and is therefore subject to stochastic fluctuations.

A deterministic solution to~(\ref{eq:eqmot}) for an arbitrarily-large number of substructures is a prohibitive task given that the trajectory of the tracer is coupled with $N-$differential equations.
As an alternative approach, Paper~I uses a statistical method originally introduced by Holtsmark (1919) to compute the probability $p({\mathbfit F})$ that a tracer particle experiences a combined force ${\mathbfit F}$ within the interval ${\mathbfit F},{\mathbfit F}+\d{\mathbfit F}$. Hernquist (1990) spheres with a density profile
\begin{align}\label{eq:rho}
  \rho(r)= \frac{M}{2\pi c^3}\frac{1}{(r/c)(1+r/c)^3},
\end{align}
induce an individual force
\begin{eqnarray}\label{eq:fh}
{\bf f}=-\frac{GM}{(r+c)^2} \hat{\bf r}.
\end{eqnarray}
The distribution of force fluctuations generated by a large population of these objects can be expressed in analytical form under the following assumptions: (i) substructures do not spatially overlap, such that their average separation ($D$) is much larger than their individual size ($c$), i.e. $c/D\ll 1$, (ii) these objects are in dynamical equilibrium within the host potential $\Phi_g$, which leads to a static distribution $p({\mathbfit F})$, (iii) they are randomly distributed within the volume $V$ or, equivalently, the forces ${\bf f}_i$ are spatially uncorrelated, and (iv) their number density can be assumed to be roughly constant, such that $n({\mathbfit r}_\star+\Delta{\mathbfit r})\approx n({\mathbfit r}_\star)=n=N/V$. This is known as the {\it local approximation}, and holds insofar as the (local) density profile rolls slowly, i.e. $d=|\nabla n/n|^{-1}\gtrsim D$ (see Appendix A of Paper~I). Under these conditions the spectrum of force fluctuations is isotropic, i.e. $p({\mathbfit F})=p(F)$. In the limit $N\to \infty$ it can be written analytically as
\begin{align}\label{eq:pF}
  p({\mathbfit F})&=\frac{1}{V}\int \d^3r_1 \times ...\times \frac{1}{V}\int \d^3r_N\,\delta\bigg({\mathbfit F}-\sum_i\,{\mathbfit f}_i\bigg) \\ \nonumber
  &\approx  \frac{C}{1+A(F/b)^{9/2}}\bigg(1-\sqrt{\frac{F}{f_0}}\bigg)^2~~~{\rm for}~~~F<f_0,
\end{align}
where $\delta$ is the Dirac's delta function; $f_0=G M/c^2$ is the maximum force generated by an individual substructure; $C$ is a normalization constant that guarantees $\int \d^3 F\,p({\mathbfit F})=1$; $A=4\pi C b^3$ is a dimension-less quantity of order unity, $b=G M/D^2$ is the average force induced by the closest object to the tracer particle, and $D=(2\pi n)^{-1/3}$ is the distance at which the probability of finding the nearest substructure peaks (see Paper~I for details). Note that in the point-mass limit $c\to 0$ ~($f_0\to \infty$) the distribution~(\ref{eq:pF}) approaches a power-law behaviour $p(F)\sim F^{-9/2}$ at large forces, $F\gg b$, as originally pointed out by Holtsmark (1919) (see also Chavanis 2009). For extended substructures the distribution $p(F)$ is truncated at $F=f_0$, which implies that the strongest fluctuation arises in the unlikely event wherein the tracer particle sits at the centre of one of those objects.

The moments of the (approximate) distribution~(\ref{eq:pF}) can be computed analytically. Let us first calculate the normalization constant $A$ using the integration variable $\xi=F/b$ and the dimension-less quantity $\chi=b/f_0=(c/D)^2\ll 1$ as
\begin{align}\label{eq:norm}
  \int \d^3 F\,p({\mathbfit F})&=4\pi \int_0^{f_0}\d F\,F^2p(F)\\ \nonumber
  &=4\pi C b^3\int_0^{1/\chi}\d \xi\,\frac{\xi^2}{1+A\xi^{9/2}}[1-\chi^{1/2}\xi^{1/2}]^2\\ \nonumber
  &= \frac{4 \pi A^{1/3}}{9\sqrt{3}}+ \mathcal{O}(\chi^{1/2}),
\end{align}
hence, using the normalization $\int \d^3 F\,p({\mathbfit F})=1$ and taking the limit $\chi=(c/D)^2\to 0$ yields $A\to [9\sqrt{3}/(4\pi)]^3\simeq 1.91$.

The average force acting on the tracer particle can be calculated as
\begin{align}\label{eq:F1}
  \langle F\rangle&= \int \d^3 F\,F\,p({\mathbfit F})\\ \nonumber
  &=4\pi C b^4\int_0^{1/\chi}\d \xi\,\frac{\xi^3}{1+A\xi^{9/2}}[1-\chi^{1/2}\xi^{1/2}]^2\\ \nonumber
  &=b\, \bigg[\frac{2\pi A^{1/9}}{9}{\rm Cosec}\bigg(\frac{\pi}{9}\bigg)  +\bigg(-\frac{4}{9}\ln A+ 2\ln\chi\bigg)\sqrt{\chi} + \mathcal{O}(\chi)\bigg].
\end{align}
Substituting $b=GM/D^2=(2\pi)^{2/3}GM n^{2/3}$ in~(\ref{eq:F1}) and taking the particle limit $\chi\to 0$ yields an average force $\langle F\rangle \simeq 2.19b\simeq 7.47 GM n^{2/3}$, which recovers Equation~(15) of Paper~I modulo a numerical factor of order unity (0.84 to be precise) that arises from the approximate nature of the distribution~(\ref{eq:pF}). It is important to notice that at leading order the average force acting on the tracer particle is independent of the size of individual substructures. We will come back to this point in \S\ref{sec:nbody}.

The variance of the force distribution is
\begin{align}\label{eq:F2}
  \langle F^2\rangle&= \int \d^3 F\,F^2\,p({\mathbfit F})\\ \nonumber
  &=4\pi C b^5\int_0^{1/\chi}\d \xi\,\frac{\xi^4}{1+A\xi^{9/2}}[1-\chi^{1/2}\xi^{1/2}]^2\\ \nonumber
  &=b^2\,\bigg[\frac{2}{3\chi}-\frac{2\pi}{9A^{1/9}}{\rm Cosec}\bigg(\frac{\pi}{9}\bigg)  +\frac{56\pi}{81 A^{2/9}}\sqrt{\chi} + \mathcal{O}(\chi)\bigg].
\end{align}
To first order in $\chi$ (which is assumed small) Equation~(\ref{eq:F2}) becomes $\langle F^2\rangle\simeq 2b^2/(3\chi)=4\pi (GM)^2n/(3c)$, thus recovering Equation~(22) of Paper~I exactly. Note that in contrast to the average force, the magnitude of the force fluctuations is very sensitive to the size of individual substructures, diverging $\langle F^2\rangle \to \infty$ in the point-mass limit $c\to 0$. In practice, this means that as time goes by the maximum force experienced by a test particle surrounded by point-masses can grow up to arbitrarily-large values, a well-known issue in computational dynamics (e.g. Heggie \& Hut 2003).

\subsection{Velocity `kicks'}\label{sec:kicks}
A test particle travelling through a clumpy medium experiences fluctuations of the local gravitational field due to the rapid change of the (relative) position of nearby objects. Chandrasekhar (1941a,b; 1943) argues that the cumulative effect of force fluctuations leads to random increments of the particle velocity, $\Delta {\mathbfit v}$, which can be treated as a random walk in a three-dimensional velocity space. For a substructure population distributed homogeneously around the tracer particle, this Brownian motion leads to a distribution of velocity impulses that is isotropic and has a Gaussian form (Chandrasekhar 1941b, 1943; Kandrup 1980)
\begin{align}\label{eq:Psi}
 \Psi({\mathbfit v}, \Delta {\mathbfit v},t)=\frac{1}{(\frac{2\pi}{3}\langle |\Delta{\mathbfit v}|^2\rangle)^{3/2}}\exp\big[-\frac{(\Delta{\mathbfit v}-\langle \Delta{\mathbfit v}\rangle)^2}{\frac{2}{3} \langle |\Delta{\mathbfit v}|^2\rangle}\big];
\end{align}
where $\Psi({\mathbfit v}, \Delta {\mathbfit v},t-t_0)$ denotes the probability that a test particle with a velocity ${\mathbfit v}$ will experience a velocity impulse $\Delta{\mathbfit v}$ within a time interval $t-t_0$. In what follows, we set $t_0=0$ for simplicity.

It is straightforward to show that the averaged velocity impulse vanishes by symmetry, $\langle \Delta{\mathbfit v}\rangle=0$, whereas the average squared velocity increment can be written as 
\begin{align}\label{eq:delv2}
   \langle |\Delta{\mathbfit v}|^2\rangle =\int_0^t\d s \int_0^t \d s'\langle {\mathbfit F}_s\cdot {\mathbfit F}_{s'}\rangle.
\end{align}
The main difficulty in solving~(\ref{eq:delv2}) is that the forces ${\mathbfit F}_s$ and ${\mathbfit F}_{s'}$ measured at two different times $s$ and $s'$ cannot be in general assumed to be statistically independent. Indeed, these two quantities are related through the trajectories of individual substructures in the host potential. However, Equation~(\ref{eq:delv2}) exhibits two well-defined asymptotic behaviours at short and long time-scales, which are inspected in some detail below.

\subsubsection{Long time-scales (dynamic limit)}\label{sec:long}
Over a sufficiently long interval of time, Equation~(\ref{eq:delv2}) can be approximated by the average squared increments of a Brownian motion in an unconfined three-dimensional velocity space (Chandrasekhar 1941b)
\begin{align}\label{eq:delv2_long}
   \langle |\Delta{\mathbfit v}|^2\rangle =t\langle F^2 T\rangle=t\int\d^3 F p({\mathbfit F})F^2 T(F),
\end{align}
where $T(F)$ is the mean life of a force fluctuation. 
This quantity can be calculated in a simple manner using Smoluchowski (1916) analysis\footnote{Chandrasekhar \& von Neumann (1942, 1943) provide an alternative, yet mathematically cumbersome method to compute $T(F)$. Kandrup (1980) shows that both approaches yield consistent results modulo a numerical factor of order unity.}, who argues that fluctuations of the local acceleration must arise from the presence (or absence) at some point of time of substructures in the vicinity of the test particle.
Consider $N$ substructures distributed within a volume $V$ around the test particle at $t=0$. At a later time, the number of substructures inside this volume will change either because one of the substructures exits or another enters from outside. According to Smoluchowski (1916), the probability $P_N(t)$ that at some later time there are still $N$ substructures inside $V$ may be written as $P_N(t)\d t =e^{-t/T} \d t/T$. Here, the time scale $T$ corresponds to the mean life of a state in which the number of substructures within $V=4\pi r^3/3$ remains constant.
For a uniform distribution of substructures with a number density $n=N/V$ and a mean squared (relative) velocity $\langle v^2\rangle$, Smoluchowski (1916) finds
\begin{align}\label{eq:TR}
  T(r)={\sqrt\frac{2\pi}{3 \langle v^2\rangle} }\frac{r}{\frac{4\pi }{3}r^3n+1}.
\end{align}
For substructures located at small distances, $4\pi r^3 n/3 = (2/3) (r/D)^3\ll 1$, Equation~(\ref{eq:TR}) converges to time that a substructure moving on a straight-line trajectory would take to cross the distance $r$, i.e. $T_{sl}(r)\sim r/\sqrt{\langle v^2\rangle}$. In contrast, at large radii the probability that a substructure leaves/enters $V$ becomes proportional to the number of substructures within this volume. Hence, in the limit $4\pi r^3 n/3 \gg 1$ Smoluchowski's timescale becomes inversely proportional to $N=n V$, such that $T\sim r/\sqrt{\langle v^2\rangle}N^{-1}\sim r^{-2}$, which vanishes in the limit $r\to \infty$.

As shown by Chandrasekhar (1941b), it is safe to assume that the force acting on a tracer particle is dominated by the nearest substructure. As shown below, this approximation is key in order to construct a theory that depends on local quantities only. For substructures with individual forces~(\ref{eq:fh}) this implies $F\approx GM/(r+c)^2$. Inserting this relation into~(\ref{eq:TR}) and re-arranging yields
\begin{align}\label{eq:TF}
  T(F)=T_0\frac{(1-\sqrt{b/f_0}\sqrt{F/b})(F/b)}{(2/3)(1-\sqrt{b/f_0}\sqrt{F/b})^3+(F/b)^{3/2}},
\end{align}
with $T_0=(2\pi/3)^{1/2} D/\langle v^2\rangle^{1/2}$. Recall that $b=GM/D^2$ and $f_0=GM/c^2$, and that our working assumption is that $b/f_0=(c/D)^2\ll 1$. For substructures moving on straight-line trajectories Equation~(\ref{eq:TF}) reduces to
\begin{align}\label{eq:TF_sl}
  T_{sl}(F)={\sqrt\frac{2\pi}{3 \langle v^2\rangle} }r=T_0\sqrt{\frac{b}{F}}\bigg(1-\sqrt{\frac{F}{f_0}}\bigg).
\end{align}

It can be readily seen from Equation~(\ref{eq:TF}) that the mean life of force fluctuations vanishes at strong ($F\to f_0$) and weak ($F\to 0$) forces, which indicates the existence of a characteristic time span, $T_{\rm ch}$, associated with the longest, and therefore most likely, force fluctuation, $F_{\rm ch}$. After some algebra, one can show that the solution to $\d T/\d F|_{F_{\rm ch}}=0$ is
\begin{align}\label{eq:Fch}
  F_{\rm ch}=\frac{2^{4/3}}{3^{2/3}}-\frac{8}{3}\frac{c}{D}+\mathcal{O}\bigg(\frac{c}{D}\bigg)^2
\end{align}
which leads to a characteristic duration of force fluctuations
\begin{align}\label{eq:Tch}
  T_{\rm ch}&=T(F_{\rm ch})=\frac{2^{1/3}}{3^{2/3}}T_0\\ \nonumber
  &\simeq 0.88 \frac{D}{\sqrt{\langle v^2\rangle}},
\end{align}
hence, $T_{\rm ch}$ roughly corresponds to the time that objects moving with an average speed $\langle v^2\rangle^{1/2}$ take to travel a distance $D$. Note that $\langle v^2\rangle$ averages over {\it relative} velocities, which explains why the distribution of velocity increments $\Psi(\mathbfit{v},\Delta\mathbfit{v},t)$ carries the velocity of the test-particle as an argument, even if this quantity does not appear explicitly in the moments of the distribution. Also, it is worth stressing that the average duration of force fluctuations $T_{\rm ch}$ does not depend on the mass or the size of the objects that cause them\footnote{Note also that the characteristic mean life of force fluctuations matches that of {\it tidal} fluctuations given by Equation~(30) of Paper~II.}.

For most cases of astrophysical interest, 
the local gravitational field fluctuates much faster than the orbital motion of the tracer particle around the host galaxy, i.e. $T_{\rm ch}\ll r/v$. This is because the typical size of a system, $r$, is typically much larger than the mean-separation between substructures, $D$, within that system.
Under this condition, it is safe to adopt the {\it impulsive approximation}, in which (i) the location of a tracer particle does not vary appreciably during a force fluctuation, and (ii) the relative motion of the closest substructure can be described by a straight line, such that $T\approx T_{sl}$. These simplifications permit an analytical solution to Equation~(\ref{eq:delv2_long}). Using~(\ref{eq:pF}) and~(\ref{eq:TF_sl}), and changing the integration variable to $\nu=A^{2/9}F/b$ yields
\begin{align}\label{eq:delv2_long_sl}
  \langle |\Delta{\mathbfit v}|^2\rangle &\approx t\int\d^3 F p({\mathbfit F})F^2 T_{sl}(F) \\ \nonumber
  &=t\,T_0 b^2\int_0^{1/\chi'}\d\nu\frac{\nu^{7/2}}{1+\nu^{9/2}}(1-\chi'^{1/2}\nu^{1/2})^3\\ \nonumber
  &=t\,T_0 b^2\bigg[\ln(1/\chi')-\frac{11}{3}+\frac{52\pi}{27}\sqrt{\chi'}+\mathcal{O}(\chi')\bigg],
\end{align}
where $\chi'=\chi/A^{2/9}=(c/D)^2/A^{2/9}$. Equation~(\ref{eq:delv2_long_sl}) has physical meaning only and only if the quantity within brackets is positive. At leading order, this implies $\ln(1/\chi')-11/3=\ln(D/c)^2-11/3+\ln A^{2/9}\gtrsim 0$, which sets a minimum distance-to-size ratio $\ln(D/c)\gtrsim 1.9$ (or $c/D\lesssim 0.15$). Here, we have used $A=1.91$ (see \S\ref{sec:sto}). For ratios $\ln(D/c)\gtrsim 1.9$, Equation~(\ref{eq:delv2_long_sl}) can be approximately written as
\begin{align}\label{eq:delv2_long_final}
  \langle |\Delta{\mathbfit v}|^2\rangle &\approx t\,\sqrt{\frac{8\pi}{3\langle v^2\rangle}}\frac{(GM)^2}{D^3}\big[\ln(D/c)-1.9\big] \\ \nonumber
    &=t\,\sqrt{\frac{32\pi^3}{3\langle v^2\rangle}}(GM)^2n\big[\ln (\Lambda) -1.9\big],
\end{align}
which exhibits a mild divergence of the so-called `Coulomb logarithm', $\ln (\Lambda)=\ln(D/c)$, in the particle limit $c\to 0$. The divergence associated with point-mass particles was originally pointed out by Chandrasekhar (1941a,b), who argued for truncating the distribution of nearby particles at a {\it critical} radius $r_{\rm crit}=2 GM/\langle v^2\rangle$.
We shall return to this issue in \S\ref{sec:disc}. In contrast, this work shows that when the same theory is applied to extended objects no {\it ad-hoc} truncations at small separations or strong forces are required. Furthermore, the Coulomb factor attains a well-defined physical meaning, $\Lambda=D/c$, which corresponds to the average distance-to-size ratio of the closest substructures.

It is also important to bear in mind that Equation~(\ref{eq:delv2_long_final}) relies on the diffusive condition $\langle |\Delta{\mathbfit v}|^2\rangle\ll v_\star^2$, where $v_\star$ represents the speed of the test particle. This condition sets an upper limit to the length of the time interval
\begin{align}\label{eq:trel}
  t\ll t_{\rm rel}\equiv\sqrt{\frac{3}{32\pi^3}} \frac{v_\star^2 \sqrt{\langle v^2\rangle}}{(GM)^2n\,[\ln(\Lambda)-1.9]}.
\end{align}
The time-scale~(\ref{eq:trel}) corresponds the {\it relaxation} time associated with tracer particles in a clumpy medium. The classical two-body relaxation time-scale of collisional, self-gravitating stellar systems is recovered by setting $M=m_\star$; $n= n_\star$ and $\langle v^2\rangle=v_\star^2$ (e.g. Spitzer 1987).

In summary, we conclude that Equation~(\ref{eq:delv2_long_final}) is only valid for substructures with individual sizes that are much smaller than their average separation, $c/D\lesssim 0.15$, and for time intervals that are sufficiently short to guarantee the validity of the diffusion equations, but long enough to permit a large number of force fluctuations, i.e. $T_{\rm ch}\ll t\ll t_{\rm rel}$.


\subsubsection{Short time-scales (static limit)}\label{sec:short}
Over very short time intervals, $t\lesssim T_{\rm ch}$, the relative position between the tracer particle and the surrounding substructures will not change appreciably with time. In this so-called {\it static} limit, velocity increments can be calculated using a familiar leap-frog integration, i.e. $\Delta{\mathbfit v} =t\,  {\mathbfit F}$. Furthermore, given that substructures are frozen in their present locations, different substructure ensembles generated at the times $s$ and $s'$ will produce statistically uncorrelated velocity increments. Hence, one can write Equation~(\ref{eq:delv2}) as $\langle|\Delta{\mathbfit v}|^2\rangle =\langle|\int_0^t\d s {\mathbfit F}_s|\rangle\langle|\int_0^t \d s' {\mathbfit F}_{s'}|\rangle=(t\langle F\rangle)^2$. Inserting the average force~(\ref{eq:F1}) and taking the leading order yields 
\begin{align}\label{eq:delv2_short}
  \langle|\Delta{\mathbfit v}|^2\rangle&= t^2\langle F\rangle^2 \\ \nonumber
  &\simeq t^2\,4.81 \,\frac{(GM)^2}{D^4}\\ \nonumber
  &\simeq t^2\,55.8 \,(GM)^2 n^{4/3}.
\end{align}
Equation~(\ref{eq:delv2_short}) implies that for very small time-intervals the average magnitude of the velocity impulses does not depend on the velocity distribution, nor the sizes of individual substructures.

Re-writing Equations~(\ref{eq:delv2_long_final}) and~(\ref{eq:delv2_short}) in terms of the density of substructures $\rho=M\,n$ shows that in a dynamic regime of fluctuations the variance of velocity impulses scales as $\langle|\Delta{\mathbfit v}|^2\rangle\sim M^2 n=M\,\rho$, whereas in a static regime
$\langle|\Delta{\mathbfit v}|^2\rangle\sim M^2 n^{4/3}=M^{2/3}\rho^{4/3}$. In both cases the {\it mean-field} limit $\langle|\Delta{\mathbfit v}|^2\rangle \to 0$ is reached as $M\to 0$ at a fixed density.

Interestingly, taking the time interval $t\to 0$ in Equation~(\ref{eq:delv2_short}) yields a characteristic acceleration 
\begin{align}\label{eq:accel}
  a=\lim_{t\to 0}\frac{\langle|\Delta{\mathbfit v}|^2\rangle^{1/2}}{t}= 2.19 \,\frac{GM}{D^2}.
\end{align}
which is largely set by the mass and average distance to the closest object.
We discuss some physical implications of this result in \S\ref{sec:sum}.

\section{$N$-body tests}\label{sec:nbody}
This Section analyzes a number of controlled N-body experiments that follow the dynamical evolution of test particles orbiting in a clumpy potential. The effect of substructures is modelled in two ways: (i) computing the random component of the gravitational force, ${\mathbfit F}$, as a direct sum of the individual forces of $N$-substructures, which is a computationally demanding --albeit exact-- approach, and (ii) by adding random velocity `kicks' to tracer particles orbiting in a smooth potential using a Monte-Carlo technique.
The numerical set-up is explained in detail in Section~5.1 of Paper~II. A brief summary is given here for completeness.
\subsection{Host and substructure models}
Tracer particles and substructures orbit in a Dehnen (1993) potential
\begin{align}\label{eq:phi_dehn}
\Phi_g(r)=\frac{4\pi G\rho_0}{3-\gamma}\times\begin{cases}
-\frac{1}{2-\gamma}\big[1-\big(\frac{r}{r+r_0}\big)^{2-\gamma}\big] & ,\gamma\ne 2 \\ 
\ln\big(\frac{r}{r+r_0}\big) & , \gamma=2.
\end{cases}
\end{align}
with a total mass $M_g=4\pi \rho_0 r_0^3/(3-\gamma)$. For simplicity, in what follows $G=\rho_0=r_0=1$. 

Substructures follow a number density profile 
\begin{align}\label{eq:n_dehn}
n(r)=\frac{n_0}{[1+(r/r_0)^\alpha]^{\beta/\alpha}},
\end{align}
with $n_0$ chosen such that $4\pi\int_0^\infty\d r\, r^2\, n(r)=N$. To assign orbital velocities in a way that guarantees dynamical equilibrium, the distribution function of substructures is calculated using Eddington (1916) inversion
\begin{align}\label{eq:eddin}
  f(E)=\frac{1}{\sqrt{8}\pi^2}\bigg[\int_E^0\frac{\d\Phi}{\sqrt{\Phi-E}}\frac{\d^2 n}{\d\Phi^2}+\frac{1}{\sqrt{-E}}\bigg(\frac{\d n}{\d \Phi}\bigg)_{\Phi=0}\bigg],
\end{align}
where $n[r(\Phi_g)]$ corresponds to the profile~(\ref{eq:n_dehn}) expressed as a function of the potential~(\ref{eq:phi_dehn}). Our models set $\gamma=0$, $\alpha=3$ and $\beta=60$, which guarantees $f(E)>0$ over the range of energies explored by these models. These parameters roughly correspond to an almost uniform
distribution of perturbers in a constant density core.
The initial radii and velocities of $N$ substructures are drawn randomly from the distribution function $f(E)$, with position and velocity vectors isotropically distributed over the surface of a sphere. The resulting velocity dispersion is $\langle v^2\rangle^{1/2}\simeq 0.45$.
To generate $N_{\rm ens}$ ensembles of substructures one must simply use different integer values in the random number generator.

\subsection{Random fluctuations}\label{sec:random}

\begin{enumerate}
\item {\bf Direct-force computation}. In this approach, the combined force induced by a set of $N$-Hernquist (1990) spheres appearing on the right-hand side of the equations of motion~(\ref{eq:eqmot}) is computed as a direct summation of the forces induced by individual substructures. From~(\ref{eq:fh}) ones has that
  \begin{align}\label{eq:tmatrix}
{\mathbfit F}= \sum_{i=1}^N \,{\mathbfit f}_i\equiv - \sum_{i=1}^N\frac{G M}{(r'_i+c)^2}\hat{\mathbfit r}'_i,
  \end{align}
  where ${\mathbfit r}'_i={\mathbfit r}_\star-{\mathbfit r}_i$ is the relative position between the tracer `star' and the $i^{th}$ substructure, and $\hat{\mathbfit r}'_i$ is a unit vector. It must be stressed that solving the equations of motion~(\ref{eq:eqmot}) with a direct-force approach becomes computationally prohibitive for $N\gtrsim 10^5$, which excludes models with realistic number of Galactic substructures.
  \vskip0.2cm
\item {\bf Monte-Carlo sampling of velocity kicks}.
  Here, the effects of a fluctuating force field are treated as a Markovian process by injecting velocity impulses at individual time-steps of the orbital evolution of tracer particles. To do so, first we compute the trajectory of a test particle in a smooth potential $\Phi_g$ at two subsquent time-steps, i.e. from $\mathbfit{r}_\star(t)\to \mathbfit{r}_\star(t+\Delta t)$. This is done by solving the equations of motion
 \begin{align}\label{eq:mc}
   \frac{{\d^2 \mathbfit r}_\star}{\d t^2}=-\nabla\Phi_g({\mathbfit r}_\star),
 \end{align}
 using a leap-frog technique.
 Next, a random velocity `kick' is added to the velocity vector computed from~(\ref{eq:mc}) as
 $${\mathbfit v}_\star(t)\to {\mathbfit v}_\star(t+\Delta t) + {\it Ran}(\Delta{\mathbfit v})$$
 where ${\it Ran}(\Delta{\mathbfit v})$ correspond to random velocity increments drawn from the isotropic Gaussian distribution $\Psi({\mathbfit v}, \Delta {\mathbfit v},\Delta t)$ given by~(\ref{eq:Psi}), and $\Delta t$ is the time-step of the leap-frog integration.

 To calculate the second moment of the probability function $\Psi$, a linear interpolation between the {\it static} and {\it dynamic} fluctuation regimes is performed as follows
\begin{align}\label{eq:delv2_int}
  \langle|\Delta{\mathbfit v}|^2\rangle= \frac{1}{1+(t/\tau_d)}\langle|\Delta{\mathbfit v}|^2\rangle_s+\frac{(t/\tau_d)}{1+(t/\tau_d)}\langle|\Delta{\mathbfit v}|^2\rangle_d,
\end{align}
where the coefficients $\langle|\Delta{\mathbfit v}|^2\rangle_d$ and $\langle|\Delta{\mathbfit v}|^2\rangle_s$ are given by Equations~(\ref{eq:delv2_long_final}) and~(\ref{eq:delv2_short}), respectively, with the length of the time-interval set to $\Delta t$. Following \S\ref{sec:short}, the transition between the two asymptotic behaviours is set on a time-scale comparable to the characteristic duration of fluctuations, $\tau_d=T_{\rm ch}/2$. As shown in \S\ref{sec:results}, this choice provides a good description of the $N$-body experiments. 
 
\end{enumerate}

The equations of motion~(\ref{eq:eqmot}) are solved for individual tracer particles and substructures using a leap-frog technique with a time-step chosen to conserve orbital energy at a $10^{-9}$ level in a smooth ($N=0$) potential. This choice guarantees that the variation of energy due to numerical errors can be neglected in the experiments shown below. 
In order to reduce computational cost, the fluctuating component of the force field is ignored in the calculation of the trajectories of individual substructures.

\begin{figure*}
\begin{center}
\includegraphics[width=170mm]{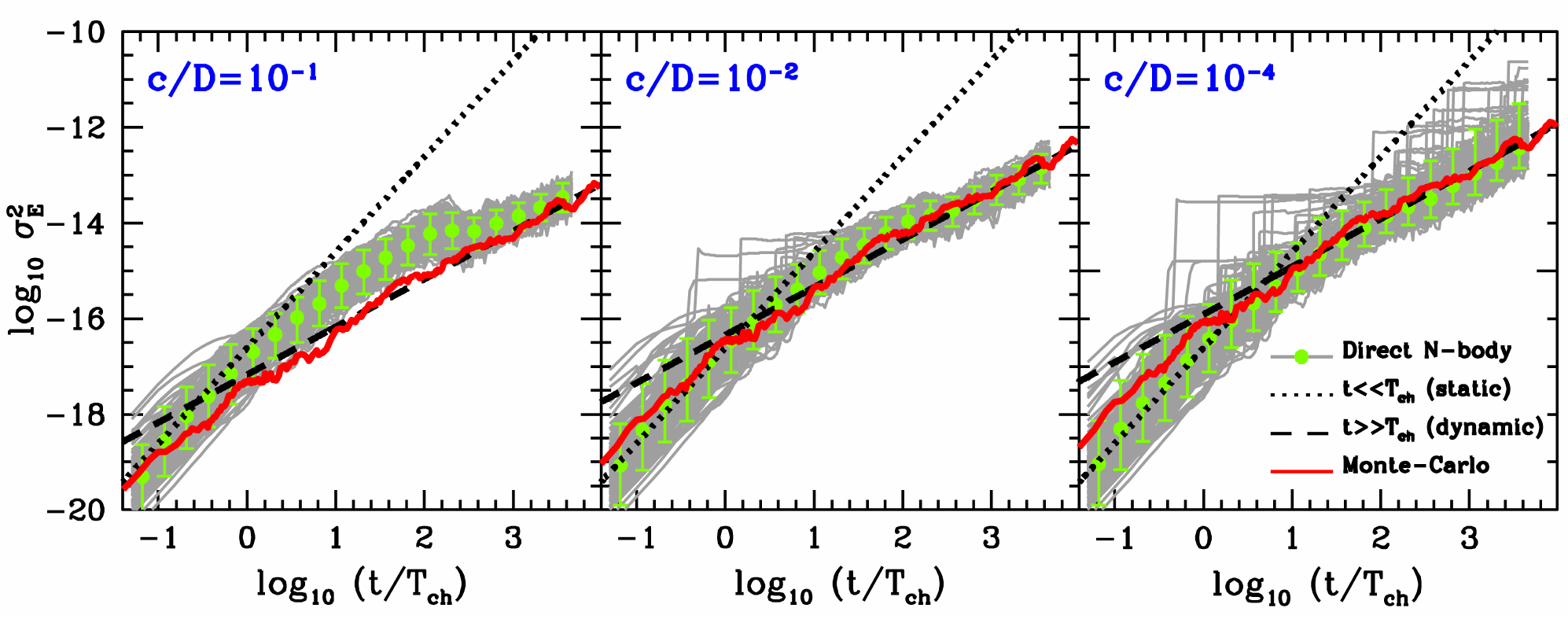}
\end{center}
\caption{Variation of orbital energy as a function of time. Grey line show the energy evolution of $N_{\rm ens}=100$ ensembles of $N_\star=20$ tracer particles acted on by stochastic forces generated by $N=10^4$ substructures with a mass $M=10^{-9}$ and a scale-radius $(c)$ given in units of the average substructure separation $(D)$. Green dots denote ensemble-averaged energy values at different time bins. Dotted and dashed black lines show the theoretical expectation in the static and dynamic limits, respectively,  Equation~(\ref{eq:delE2}). Monte-Carlo sampling velocity kicks (red lines) with the interpolation~(\ref{eq:delv2_int}) provides a reasonable match to the average energy evolution obtained from direct $N$-body models on short ($t/T_{\rm ch}\lesssim 1$) and long ($t/T_{\rm ch}\gg 1$) time-intervals.}
\label{fig:de_3}
\end{figure*}

\subsection{Test particle set-up}\label{sec:setup}
A convenient approach to test the analytical equations of \S\ref{sec:kicks} is to inspect the evolution of energy $E$ and angular momentum vector ${\mathbfit L}$ of test particles, as these quantities are conserved in the absence of substructures. Following Paper~II, let us write the average variation of energy and angular momentum due to isotropic velocity impulses as
\begin{align}\label{eq:delEL}
  \langle \Delta E\rangle &=\langle{\mathbfit v}_\star\cdot\Delta{\mathbfit v}\rangle+ \frac{1}{2}\langle|\Delta {\mathbfit v}|^2\rangle=\frac{1}{2}\langle|\Delta {\mathbfit v}|^2\rangle \\ \nonumber 
   \langle \Delta {\mathbfit L}\rangle &=\langle{\mathbfit r}_\star\times\Delta {\mathbfit v}\rangle=0.
\end{align}
where $\Delta E=E-E_0$ and $\Delta {\mathbfit L}= {\mathbfit L}-{\mathbfit L}_0$, with $E_0=E(t=0)$ and ${\mathbfit L}_0={\mathbfit L}(t=0)$.
In this notation, brackets denote averages over the spectrum of force fluctuations, i.e. $\langle X\rangle=\int \d^3 F\, p({\mathbfit F}) \,X$.

Given the significant computational cost of direct-force calculations, it is helpful to run numerical experiments that are minimally dependent on the number of substructures in the simulation. With this aim in mind, we place test particles on circular orbits, $v_\star(t=0)=v_c(r_\star)=[GM_g r_\star^{2-\gamma}/(r_\star+r_0)^{3-\gamma}]^{1/2}$, at galactocentric radius that is equal to the average separation between substructures, $r_\star(t=0)=D$, with angular momentum vectors that point in random directions over the surface of a sphere.
In a cored potential~(\ref{eq:phi_dehn}) with $\gamma=0$, the choice of initial conditions leads to a circular velocity $v_c^2(D)=G M_g D^2/(D+r_0)^3\simeq GM_g D^2/r_0^3$, where it is assumed that the relative distance between substructures is much smaller than the scale radius of the host galaxy, i.e. $ D\ll r_0$, which typically implies $N\gg 1$.

Using the initial conditions ${r_\star^2}=D^2\gg c^2$ and ${v_\star^2}=GM_g D^2/r_0^3$, and computing the variance of the integrals at leading order $\mathcal{O}(|\Delta {\mathbfit v}|/v_\star)$ yields (see Paper~II for details)
\begin{align}\label{eq:delE2}
  \sigma_E^2(t)&= \overline{\langle (\Delta E)^2\rangle}-\overline{\langle \Delta E \rangle^2}\\ \nonumber
 & \simeq  \frac{1}{3}\overline{v_\star^2}\langle |\Delta{\mathbfit v}|^2\rangle \\ \nonumber
  &\approx\frac{G M_g}{r_0^3}\frac{(GM)^2}{\langle v^2\rangle}\times\begin{cases}
1.24\,(t/T_{\rm ch})^2 & ,t\ll T_{\rm ch} \\ 
0.85\,(t/T_{\rm ch})\,[\ln(D/c)-1.9] & , t\gg T_{\rm ch}.
\end{cases}
\end{align}
and
\begin{align}\label{eq:delL2}
  \sigma_L^2(t)&= \overline{\langle|\Delta {\mathbfit L}|^2\rangle}-\overline{\langle\Delta {\mathbfit L}\rangle^2}\\ \nonumber
  &= \frac{2}{3}\overline{r_\star^2}\,\langle |\Delta{\mathbfit v}|^2\rangle \\ \nonumber
  &\approx\frac{(GM)^2}{\langle v^2\rangle}\times\begin{cases}
2.48\,(t/T_{\rm ch})^2 & ,t\ll T_{\rm ch} \\ 
1.70\,(t/T_{\rm ch})\,[\ln(D/c)-1.9] & , t\gg T_{\rm ch},
\end{cases}
\end{align}
where $T_{\rm ch}$ is the characteristic duration of a force fluctuation~(\ref{eq:Tch}), and upper bars $\overline{X}$ denote averages of the quantity $X$ over ensembles of tracer particles.
As desired, on short time-scales $t/ T_{\rm ch}\ll 1$ Equations~(\ref{eq:delE2}) and~(\ref{eq:delL2}) do not depend explicitly on the number density of substructures. However, for longer time-intervals, $t/ T_{\rm ch}\gg 1$, both $\sigma_E^2$ and $\sigma_L^2$ show a weak dependence on the number of background objects via the Coulomb logarithm, $\ln (D/c)$, given that $D$ scales as $D=(2\pi n)^{-1/3}\sim N^{-1/3}$. As a simple remedy, the $N$-body experiments shown below are run with fixed distance-to-size ratios.

\subsection{Results}\label{sec:results}
The models presented in this Section aim to probe conditions in which the analytical Equations~(\ref{eq:delE2}) and~(\ref{eq:delL2}) are expected to fail. In particular, our numerical experiments explore cases where the number of subhaloes is small ($N\to 1$), and the substructure size approaches the particle limit $c/D\to 0$. 
Due to the limited number of substructures that can be followed with direct-force calculations, it is important to quantify the scatter introduced by the finite sampling of the distribution function~(\ref{eq:eddin}). This is done by averaging quantities of interest over a relatively large number of random ensembles of substructures ($N_{\rm ens})$ and tracer particles ($N_\star$). 

Fig.~\ref{fig:de_3} plots the variance of energy variations ($\sigma_E^2$) as a function of time obtained from the ensemble average of $N_\star=20$ tracer particles with fixed energy and angular momentum at $t=0$ subject to the combined force induced by $N=10^4$ Hernquist spheres with a mass $M=10^{-9}$, and a scale radius $c$ that varies from $c/D=10^{-1}$ (left panel) down to $c/D=10^{-4}$ (right panel). 
The range of size-to-distance ratios is representative of those found in cold dark matter substructures of Milky Way-like haloes (see Fig.~10 of paper II).
Grey lines correspond to individual substructure ensembles. Green dots with error bars plot the median and the 10\% and 90\% probability intervals derived from $N_{\rm ens}=500$ ensembles. The average variation of angular momentum as a function of time, $\sigma_L^2$(t), has a similar behaviour, and is not shown here for brevity.

Comparison between the $N$-body models and the analytical formulae~(\ref{eq:delE2}) shows that the stochastic framework 
provides an excellent match to the numerical experiments for time-intervals that are either much shorter ($t\lesssim T_{\rm ch}$, `static limit') or much longer ($t\gg T_{\rm ch}$, `dynamic limit') than the characteristic duration of force fluctuations. Because these formulae are derived for non-overlapping substructures ($c/D\ll 1$), the discrepancy betweeen the numerical values of $\sigma_E^2$ and the theoretical dashed lines is systematically smaller as the size-to-distance ratio shrinks.
For example, extended models with $c/D=10^{-1}$ converge towards the theoretical curves on a time-scale $t\sim 10^3\, T_{\rm ch}$, whereas this time shortens down to $t\sim 10^2\, T_{\rm ch}$ for models with $c/D=10^{-2}$.
Note also that during very short time-intervals $t\lesssim T_{\rm ch}$ the variation of energy is insensitive to the size of substructures, as predicted in \S\ref{sec:short}.

Remarkably, sampling velocity kicks from the Gaussian distribution $\Psi({\mathbfit v},\Delta {\mathbfit v},t)$, Equation~(\ref{eq:Psi}), leads to energy variations (solid red lines) that closely follow the median evolution of the $N$-body models. Notice also the good agreement betweeen Monte-Carlo and direct-force models at $t\ll T_{\rm ch}$ and $t\gg T_{\rm ch}$.
Given that the computational cost of our direct-force experiments scales with $\sim N$, it is worth stressing that the Monte-Carlo technique introduced in \S\ref{sec:random} speeds up the orbital integration by many orders of magnitude.

\begin{figure*}
\begin{center}
\includegraphics[width=170mm]{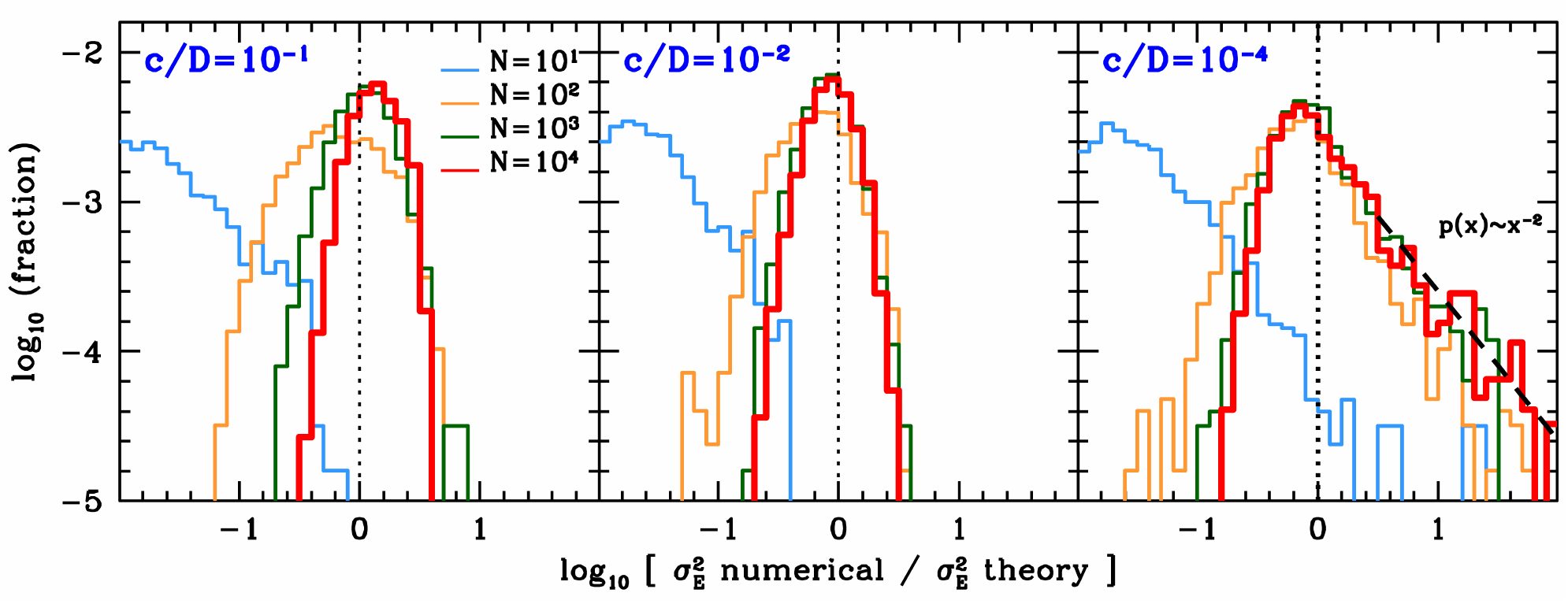}
\end{center}
\caption{Ratio between the average energy variance ($\sigma_E^2$) measured from direct-force calculations the theoretical value predicted by Equation~(\ref{eq:delE2}). The distribution uses the energies from $N_{\rm ens}=500$ random substructure ensembles measured at the final time bin of Fig.~\ref{fig:de_3}, $t/T_{\rm ch}=5000$. Full lines are coloured coded according to the number of substructures in each ensemble ($N$). Numerical convergence is attained for ensembles with $N\gtrsim 100$. For extended substructures ($c/D\gtrsim 10^{-2}$) the resulting energy distribution that has a Gaussian shape. However, close encounters with compact substructures ($c/D\ll 10^{-2}$) lead to the formation of a power-law, high-energy tail that scales as $p(x)\sim x^{-2}$, with $x=\sigma_{E,{\rm numerical}}^2/\sigma_{E,{\rm theoretical}}^2$ (marked with black dashed line for reference) and cannot be reproduced by random walk models.}
\label{fig:dist_3}
\end{figure*}

Comparison between the left and right panels of Fig.~\ref{fig:de_3} shows that the scatter around the median values increases as the Hernquist spheres become more compact in relation to their average separation. To inspect this issue in more detail, Fig.~\ref{fig:dist_3} plots the energy distribution of the tracer particles measured at the final integration time of the experiments, $t/ T_{\rm ch}=5000$. Notice first that the energy distribution converges for populations containing $N\gtrsim 10^2$ objects, and that this result holds independently of the value of the size-to-distance ratio. This confirms the numerical set-up presented in \S\ref{sec:setup}, which was specifically designed to avoid an explicit dependence on the number density of substructures. As the $N$ increases the distribution becomes narrower, which follows from the central limit theorem.
However, when the number of objects falls down to $N\sim 10$ (blue lines), Equation~(\ref{eq:delE2}) systematically overestimates the energy variation.
This mismatch is to be expected given that the force distribution $p({\mathbfit F})$ was derived in \S\ref{sec:sto} in the limit $N\to \infty$. Our numerical experiments indicate that the stochastic formalism outlined in this paper cannot be applied to populations that contain less that $N\lesssim 100$ members.

The energy distribution of relatively extended substructures ($c/D\gtrsim 10^{-2}$) has a Gaussian shape with a median value approximately centred at the energy predicted by Equation~(\ref{eq:delE2}) (marked with dotted vertical lines). 
In contrast, substructures with very small size-to-distance ratios ($c/D\lesssim 10^{-4}$) 
exhibit a power-law tail at high energies that scales as
$p(x)\sim 1/x^2$, with $x=\sigma^2_{E,{\rm numerical}}/\sigma^2_{E,{\rm theoretical}}$
(plotted with a black-dashed line for reference). This tail is due to very close encounters with low probability that cannot be treated by the random walk theory presented in \S\ref{sec:sto}, which relies on the condition $|\Delta E|\ll |E|$.
In the limit where the size of individual substructures becomes negligible with respect to their average separation ($c/D\approx 0$), one must instead use Rutherford two-body scattering. In this statistical framework, the probability that a tracer particle with an energy $E$ changes its binding energy to $E+\sigma_E$ in a short time interval $t$ has as a power-law form
 $p(\sigma_E)\sim t\,(GM)^2n/(\sqrt{\langle v^2\rangle}\sigma_E^{3})$ (Goodman 1983).
The high-energy tail marked in Fig.~\ref{fig:dist_3} with a black-dashed line scales as $p(\sigma_E^2)\sim 1/(\sigma_E^2)^2$. Applying the chain rule yields $p(\sigma_E)=2\sigma_E\,p(\sigma_E^2)\sim 1/\sigma_E^3$, which is consistent with the power-law at high energies arising from two-body scattering interactions.
Given that point-mass objects $c/D\to 0$ bring the maximum $\sigma_E^2$ to arbitrarily-large values, the ensemble average energy increment diverges logarithmically as $\overline{\sigma_E^2}=\int_0^{\infty}\d x\,p(x)\,x \propto \int_0^{\infty}\d x \,(1/x) \to \infty$. This issue is discussed in detail below.

\section{Discussion: the Coulomb logarithm}\label{sec:disc}
Chandrasekhar (1941a) was the first to point out that tracer stars moving in an homogeneous sea of point-mass particles experience two-body encounters that lead to {\it logarithmically divergent} velocity impulses when the impact parameter is taken to infinitely small values. This serious drawback was attributed firstly, to the general overestimation of the kinetic energy exchange $\Delta E$ given by two-body scattering for arbitrarily close distances and, secondly, to the increasing difficulty of describing two-body encounters as independent events when the corresponding impact parameter becomes of the same order as the intra-particle distance, $D$. He then concludes that ``{\it a consideration of this and other related difficulties suggests that we abandon the two-body approximation of stellar encounters altogether and devise a more satisfactory statistical method''}. To this aim, Chandrasekhar (1941b) applies a method originally devised by Holtsmark (1919) to describe the motion of charged particles in a plasma, which unfortunately also returns divergent velocity increments when the range of forces is taken to infinity. As a remedy, a minor modification of Holtsmark's method is proposed: ``{\it... given that the highest fields are produced by the nearest neighbor and, further, that the lack of randomness becomes significant only as $r\ll D$, it appears that we may incorporate the main features ... by supposing that no star has a first neighbor closer than $r_{\rm crit} = 2GM/\langle v^2\rangle$ and that the distribution is random but for this restriction.}''. Here, $r_{\rm crit}$ is the so-called {\it critical} radius, which is defined as the distance at which a tracer star becomes energetically bound to the nearest particle.
The resulting velocity impulses averaged over the spectrum of force fluctuations scale as $\langle \Delta{\mathbfit v}^2\rangle \propto \ln(\Lambda)$, where $\Lambda=D/r_{\rm crit}$ is typically known as the Coulomb factor\footnote{It should be noted that Chandrasekhar calculates the intra-particle separation as $D'=(4\pi n/3)^{-1/3}$, which is a factor $(3/2)^{1/3}$ larger than the value used here, $D=(2\pi n )^{-1/3}$. } in analogy with a similar quantity appearing in plasma dynamics. Numerical experiments in \S\ref{sec:results} confirm that the Brownian theory of force fluctuations breaks down below some small scale-length owing to the singular field generated by point-masses at $r=0$. However, a truncation of the distribution of nearby particles at $r_{\rm crit}$ still seems an arbitrary choice. Crucially, it is a choice that introduces mass and relative velocity dependencies into the Coulomb logarithm.

This work shows that when Chandrasekhar's (1941b) theory is applied to {\it extended} objects with a non-vanishing scale-length ($c>0$), the Coulomb factor attains a well-defined geometrical meaning, $\Lambda=D/c$, which corresponds to the average distance-to-size ratio of the nearest substructures (see Section~\ref{sec:long}). As expected, the divergence originally pointed out by Chandrasekhar re-appears for point-masses, $c= 0$. The question arises as to how the new theory behaves in the intermediate stages, that is in the limit $c/D\to 0$ where substructures have a very small (but not zero) size relative to their average separation,. 

\begin{figure}
\begin{center}
\includegraphics[width=84mm]{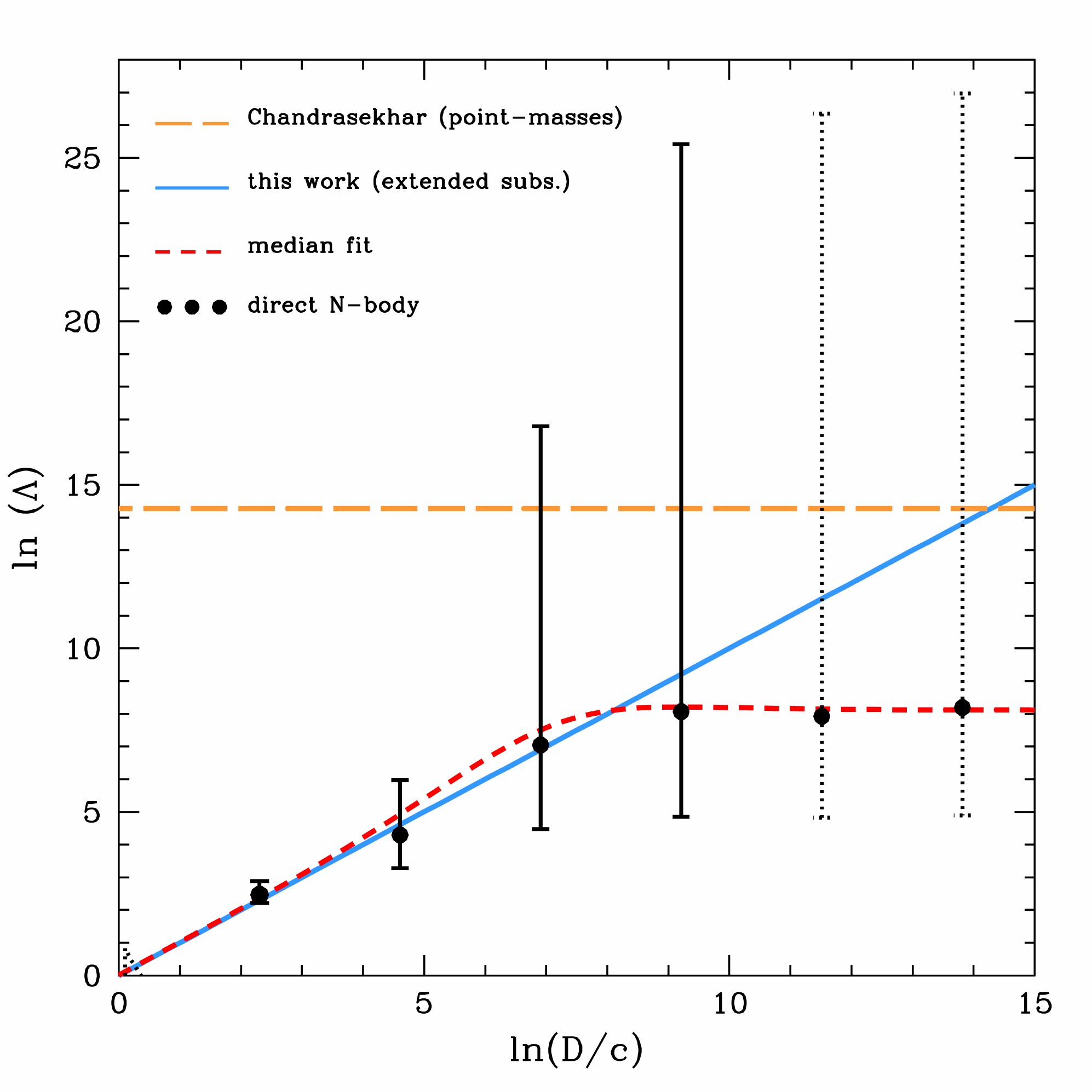}
\end{center}
\caption{Effective Coulomb logarithm derived from fitting the energy variation $\sigma_E^2(t)$ with Equation~(\ref{eq:delE2}) as a function of the distance-to-size ratio of extended substructures, $D/c$. Solid dots denote median values of the ensemble distribution, whereas the error bars mark 10\% and 90\% confidence intervals. Dotted error bars denote models for which sampling of strong forces is incomplete (see text).
  Blue curve denotes the theoretical expectation, $\ln (\Lambda)=\ln(D/c)$, whereas the dashed-orange curve plots Chandrasekhar (1941)'s Coulomb logarithm for point-mass objects, $\ln(\Lambda)=\ln(D/r_{\rm crit})$, where $r_{\rm crit}=2GM/\langle v^2\rangle$ is the critical radius. Note that in the particle limit $c\to 0$ the median value plateaus at $\ln(\Lambda)\approx 8.2$, while the upper limit keeps rising owing to the large velocity impulses induced by close encounters.}
\label{fig:lnL}
\end{figure}

To clarify this issue, it is helpful to measure the `effective' Coulomb logarithm that replaces $\ln(D/c)$ in Equation~(\ref{eq:delE2})
such that it reproduces the value of $\sigma_E^2(t)$ measured from direct-force calculations. Fig.~\ref{fig:lnL} shows how the effective Coulomb logarithm varies as a function of the size-to-distance ratio $c/D$ at a fixed time $t/T_{\rm ch}=5000$, which is sufficiently long as to guarantee a random walk behaviour $\sigma_E^2(t)\propto t$ of the $N$-body models (see Fig.~\ref{fig:de_3}). Solid dots denote the median of the distribution, whereas lower and upper error bars indicate 10\% and 90\% confidence intervals, respectively. 

As substructures become more compact, the distribution of $\ln(\Lambda)$ shows a remarkable increase of scatter at large values, which originates from the power-law tail at high energies highlighted in the right panel of Fig.~\ref{fig:dist_3}. Note that the error bars of the two most-compact models with $c/D=10^{-5}$ and $c/D=10^{-6}$ cover a similar range of values. This is likely a numerical artefact due to the limited number of random ensembles used to compute the distribution. 
Indeed, \S\ref{sec:fluctu} shows that sampling strong forces ($F\sim f_0$) requires impact parameters comparable to the substructure size. Given that the probability of close encounters scales as $p(r)\sim r^{2}$ at distances $r\ll D$ (see \S~5.2 of Paper II), the number of ensembles needed to properly sample the distribution $p({\mathbfit F})$ at strong forces goes as $N_{\rm ens}\sim (D/c)^2$. Hence, an accurate sampling of the energy distribution for objects with $c\sim r_{\rm crit}=2GM/\langle v^2\rangle=6.3\times 10^{-7}D$, requires a number of random ensembles $N_{\rm ens}$ a factor $\sim 10^4$ larger than the number considered in this work, which is computationally unfeasible.

Comparison with the analytical expectation $\ln(\Lambda)=\ln(D/c)$ (blue solid line) shows excellent agreement for relatively extended substructures ($c/D\gtrsim 10^{-3}$). 
However, as the distance-to-size ratio increases, systematic deviations from this formula becomes visible. In particular, models that approach the particle limit $c/D\to 0$ exhibit a median value of the Coulomb logarithm that plateaus at $\langle \ln(\Lambda)\rangle\approx 8.2$. The associated Coulomb factor corresponds to a minimum substructure size-to-distance ratio $c_{\rm min}/D\approx 3\times 10^{-4}$.
The convergent behaviour of $\langle \ln(\Lambda)\rangle$ is due to the fact that close encounters with compact objects extend the power-law tail in the right panel of Fig.~\ref{fig:dist_3} towards higher values, but have little effect on the median value of the distribution.
The variation of the median value of the Coulomb logarithm as a function of substructure size in Fig.~\ref{fig:lnL} can be empirically described with a linear interpolation function (red-dashed line) 
 \begin{align}\label{eq:lnL}
    \langle \ln(\Lambda)\rangle(x)=\ln(x)\frac{ x/x_{\rm int}}{1+x/x_{\rm int}} + \ln(x_{\rm max})\frac{1}{1+x/x_{\rm int}},
 \end{align}
 where $x=D/c$ is the distance-to-size ratio, $x_{\rm int}\approx 10^{-3}$ is an interpolation variable, and $\ln(x_{\rm max})\approx 8.2$. The parameter $x_{\rm int}$ determines the ratio that divides the behaviour between extended objects and point-masses in Chandrasekhar's theory of force fluctuations.
In particular, relatively extended substructures ($c/D\gtrsim x_{\rm int}$) induce a large number of small velocity impulses, whose cumulative effect can be well described by a diffusion process in velocity space (see \S\ref{sec:kicks}). In contrast, close encounters with compact substructures ($c/D\lesssim x_{\rm int}$) are governed by Rutherford's two-body scattering (see \S\ref{sec:results}). Such close encounters are, however, very rare and do not shift the median value of the energy distribution significantly.

The `truncation' of the substructure size at $c_{\rm min}\approx 3\times 10^{-4}D$ is approximately three orders of magnitude larger than the critical radius of our substructure models, $r_{\rm crit}\approx 6.3\times 10^{-7}D$. As a result, Chandrasekhar (1941b)'s formula for the Coulomb logarithm lies systematically above Equation~(\ref{eq:lnL}), $\ln(D/r_{\rm crit})\approx 14.3$ (orange-dashed line).
It is important to stress that the truncation of the distribution of nearby particles proposed by Chandrasekhar (1941b) depends on the mass and relative velocity of nearby substructures. To test this prediction, we run additional models with masses $M=10^{-10}$ and $10^{-11}$ (not shown here for brevity). These models have critical radii $r_{\rm crit}=6.3\times 10^{-8}D$ and $6.3\times 10^{-9}D$, which yield $\ln(D/r_{\rm crit})=16.6$ and 18.9, respectively. In contrast, the median measured from the $N$-body models consistently plateaus at $\langle \ln(\Lambda)\rangle\approx 8.2$ independently of the value of $r_{\rm crit}$, hence providing empirical evidence that the effective truncation $c_{\rm min}$ does not depend on the mass (or relative velocity) of nearby substructures.
  
\section{Summary}\label{sec:sum}
This paper uses direct-force calculations and Monte-Carlo techniques to study the scattering of tracer particles moving in a clumpy potential with a large number of {\it extended} substructures in dynamical equilibrium.
Following up the work of Chandrasekhar (1941b), a theory of Brownian motion is constructed which treats impulsive orbital scattering as a random walk in an unconfined three-dimensional velocity space, thus reducing the problem to the computation of drift and diffusion coefficients $\langle \Delta{\mathbfit v}\rangle$ and $\langle |\Delta{\mathbfit v}|^2\rangle$, respectively. To perform this computation the fluctuating part of the gravitational field acting on a tracer particle is described in terms of two functions: a probability density $p({\mathbfit F})$, which determines the probability of occurrence of a fluctuation ${\mathbfit F}=\sum_{i=1}^N{\mathbfit f}_i$, where ${\mathbfit f}_i$ is the force induced by a single substructure, and a function $T(F)$ which gives the average duration of a force fluctuation $F$. 
Here, the results of Paper I and II are applied, which derive an analytical expression for the probability function $p({\mathbfit F})$ associated with a large population of Hernquist (1990) spheres following a method originally devised by Holtsmark (1919), and compute the mean-life of the fluctuation $F$ using a formula due to Smoluchowski (1916).

Comparison with direct-force $N$-body experiments shows that the Brownian theory provides an excellent description of the variation of the integrals of motion during time-intervals that are either much shorter ($t\ll T_{\rm ch}$, `static' limit) or much longer ($t\gg T_{\rm ch}$, `dynamic limit) than the average duration of a fluctuation, $T_{\rm ch}\approx 0.88 D/\sqrt{\langle v^2\rangle}$, where $\sqrt{\langle v^2\rangle}$ is the average velocity between a tracer particle and a population of substructures separated by an average distance $D$.
However, the theory fails in the point-mass limit $c/D\to 0$, that is when the scale-length of individual Hernquist spheres $c$ becomes negligible with respect to their average separation. The reason behind the mismatch can be traced back to strong kinetic impulses that a tracer particle receives during low-probability close encounters with compact objects, which cannot be treated within a diffusion framework. Specifically, direct-force calculations in Fig.~\ref{fig:dist_3} show that close encounters produce a power-law tail of large velocity kicks that deviates from the Gaussian distribution predicted by the central limit theorem (see also Lee 1968).

Chandrasekhar (1941b) showed that the singular behaviour of the point-mass force at $r=0$ leads to a diffusion coefficient that diverges logarithmically. As a remedy, he introduced an ad-hoc truncation of the distribution of nearby substructures at the critical radius $r_{\rm crit}=2GM/\langle v^2\rangle$, which leads to a diffusion coefficient that scales as $\langle |\Delta{\mathbfit v}|^2\rangle\propto \ln(\Lambda)$, where $\ln(\Lambda)=\ln(D/r_{\rm crit})$ is the so-called Coulomb logarithm.
Crucially, this paper shows that when the Brownian motion theory is applied to extended objects no ad-hoc truncations at small separations or strong forces are required. Furthermore, the Coulomb factor acquires a well-defined physical meaning, $\Lambda=D/c$, which corresponds to the average distance-to-size ratio of the closest substructures. 
This simple analytical formula provides an excellent description of the empirical values measured from $N$-body experiments in which substructures have size-to-distance ratios within $10^{-1}\gtrsim c/D\gtrsim 10^{-3}$.  Interestingly, for point-like substructures ($c/D\ll 10^{-3}$) the median value of the Coulomb logarithm deviates from the logarithmic relation, flattening at a maximum value $\langle \ln(\Lambda)\rangle \approx 8.2$ that is independent of the critical radius of the models.
This behaviour is described in Section~\ref{sec:disc} with an empirical interpolation, which in principle can be used to extend the Brownian motion formalism to objects with arbitrary sizes. In practice, however, it would be desirable to construct a follow-up theory that describes how the ensemble distribution of $\ln(\Lambda)$ varies as a function of the relative size and distance of the substructure population. This goal goes beyond the diffusion models adopted here, and requires a full statistical treatment of the `heavy-tailed', non-Gaussian distribution of energy impulses generated by close encounters with point-like objects (e.g. by treating these events as an anomalous diffusion process, Bar-Or et al. 2013).

Section~\ref{sec:random} presents a simple Monte-Carlo technique that mimics the effect of a fluctuating field by drawing random velocity impulses from a Gaussian distribution~(\ref{eq:Psi}) and adding them to the orbital velocity at individual time-steps. The resulting orbital evolution agrees well with the median variation of the integrals of motion derived from direct $N$-body models (see Fig.~\ref{fig:dist_3}), which opens up the possibility to model the dynamics of tracer populations acted on by an arbitrarily-large number of substructures at a minimum computational cost. 
Future applications of this method may include, for example, (i) modelling the scattering of stellar streams due to dark subhaloes (Ibata et al. 2002; Johnston et al. 2002; Yoon et al. 2011; Carlberg 2012, 2013; Erkal \& Belokurov 2015; Erkal et al. 2016; Ngan et al. 2016, Bovy et al. 2017), and baryonic substructures (Amorisco et al. 2016; Bonaca et al. 2018); (ii) exploring the survival of stellar pairs in the Milky Way disc to interactions with stars and Giant Molecular Clouds (Kamdar 2019), (iii) investigating the effect of MACHOS and Primordial Black Holes on the collisional relaxation of ultra-faint dwarf spheroidals (Brandt et al. 2016; Koushiappas \& Loeb 2017), and (iv) determining the contribution of Giant Molecular Clouds and dark matter subhaloes to disc heating and radial migration in the Milky Way (e.g. Aumer et al. 2017).
The Monte-Carlo technique also describes the acceleration of tracer particles in a clumpy environment over arbitrarily-short time intervals, thus providing an efficient statistical tool to probe the sensitivity of pulsar timing arrays to primordial black holes and dark matter microhaloes (Siegel et al. 2007; Baghram et al. 2011; Kashiyama \& Oguri 2018; Dror et al. 2019).

\section{Acknowledgements}
It is a pleasure to thank Mike Petersen, Mark Gieles and Eugene Vasiliev for useful comments \& suggestions, as well as to the anonymous referee for a careful examination of the manuscript and a constructive report.


\begin{thebibliography}{}
  
\bibitem[Amorisco et al.(2016)]{2016MNRAS.463L..17A} Amorisco, N.~C., G{\'o}mez, F.~A., Vegetti, S., \& White, S.~D.~M.\ 2016, \mnras, 463, L17 

\bibitem[Aumer et al.(2017)]{2017MNRAS.470.3685A} Aumer, M., Binney, J., \& Sch{\"o}nrich, R.\ 2017, \mnras, 470, 3685
  
  \bibitem[\protect\citeauthoryear{Baghram, Afshordi \& Zurek}{2011}]{2011PhRvD..84d3511B} Baghram S., Afshordi N., Zurek K.~M., 2011, PhRvD, 84, 043511

  \bibitem[Bar-Or et al.(2013)]{2013ApJ...764...52B} Bar-Or, B., Kupi, G., \& Alexander, T.\ 2013, \apj, 764, 52
    
 
  \bibitem[Bonaca et al.(2018)]{2018arXiv181103631B} Bonaca, A., Hogg, D.~W., Price-Whelan, A.~M., \& Conroy, C.\ 2018, arXiv:1811.03631
    
  \bibitem[Bovy et al.(2017)]{2017MNRAS.466..628B} Bovy, J., Erkal, D., \& Sanders, J.~L.\ 2017, \mnras, 466, 628

  \bibitem[Brandt(2016)]{2016ApJ...824L..31B} Brandt, T.~D.\ 2016, \apjl, 824, L31
    
\bibitem[Carlberg(2012)]{2012ApJ...748...20C} Carlberg, R.~G.\ 2012, \apj, 748, 20 

\bibitem[Carlberg(2013)]{2013ApJ...775...90C} Carlberg, R.~G.\ 2013, \apj, 775, 90
  
 \bibitem[Chandrasekhar(1941)]{1941ApJ....93..285C} Chandrasekhar, I.~S.\ 1941a, \apj, 93, 285 

  
\bibitem[Chandrasekhar(1941)]{1941ApJ....94..511C} Chandrasekhar, S.\ 1941b, \apj, 94, 511 

\bibitem[Chandrasekhar \& von Neumann(1942)]{1942ApJ....95..489C} Chandrasekhar, S., \& von Neumann, J.\ 1942, \apj, 95, 489 


\bibitem[Chandrasekhar \& von Neumann(1943)]{1943ApJ....97....1C} Chandrasekhar, S., \& von Neumann, J.\ 1943, \apj, 97, 1 

\bibitem[Chandrasekhar(1943)]{1943RvMP...15....1C} Chandrasekhar, S.\ 1943, Reviews of Modern Physics, 15, 1

  
\bibitem[Chavanis(2009)]{2009EPJB...70..413C} Chavanis, P.~H.\ 2009, European Physical Journal B, 70, 413

\bibitem[Cohen et al.(1950)]{1950PhRv...80..230C} Cohen, R.~S., Spitzer, L., \& Routly, P.~M.\ 1950, Physical Review, 80, 230

\bibitem[Colpi et al.(1999)]{1999ApJ...525..720C} Colpi, M., Mayer, L., \& Governato, F.\ 1999, \apj, 525, 720
  
\bibitem[Dehnen(1993)]{1993MNRAS.265..250D} Dehnen, W.\ 1993, \mnras, 265, 250

\bibitem[Diemand et al.(2005)]{2005Natur.433..389D} Diemand, J., Moore, B., \& Stadel, J.\ 2005, \nat, 433, 389 

\bibitem[Dror et al.(2019)]{2019arXiv190104490D} Dror, J.~A., Ramani, H., Trickle, T., \& Zurek, K.~M.\ 2019, arXiv:1901.04490
  
\bibitem[Eddington(1916)]{1916MNRAS..76..572E} Eddington, A.~S.\ 1916, \mnras, 76, 572

\bibitem[Erkal \& Belokurov(2015)]{2015MNRAS.454.3542E} Erkal, D., \& Belokurov, V.\ 2015, \mnras, 454, 3542 
  
\bibitem[Erkal et al.(2016)]{2016MNRAS.463..102E} Erkal, D., Belokurov, V., Bovy, J., \& Sanders, J.~L.\ 2016, \mnras, 463, 102 

\bibitem[Errani \& Pe{\~n}arrubia(2019)]{2019arXiv190601642E} Errani, R., \& Pe{\~n}arrubia, J.\ 2019, arXiv:1906.01642

\bibitem[Fouvry \& Bar-Or(2018)]{2018MNRAS.481.4566F} Fouvry, J.-B., \& Bar-Or, B.\ 2018, \mnras, 481, 4566
  

\bibitem[Goodman(1983)]{1983ApJ...270..700G} Goodman, J.\ 1983, \apj, 270, 700

\bibitem[Hamilton et al.(2018)]{2018MNRAS.481.2041H} Hamilton, C., Fouvry, J.-B., Binney, J., \& Pichon, C.\ 2018, \mnras, 481, 2041
  
\bibitem[Heggie \& Hut(2003)]{2003gmbp.book.....H} Heggie, D., \& Hut, P.\ 2003, The Gravitational Million-Body Problem: A Multidisciplinary Approach to Star Cluster Dynamics, by Douglas Heggie and Piet Hut.~ Cambridge University Press, 2003, 372 pp.,  



\bibitem[Hernquist(1990)]{1990ApJ...356..359H} Hernquist, L.\ 1990, \apj, 356, 359
 
  
\bibitem[Hofmann et al.(2001)]{2001PhRvD..64h3507H} Hofmann, S., Schwarz, D.~J., \& St{\"o}cker, H.\ 2001, \prd, 64, 083507 

  
\bibitem[Holtsmark(1919)]{1919AnP...363..577H} Holtsmark, J.\ 1919, Annalen der Physik, 363, 577 

  

\bibitem[Ibata et al.(2002)]{2002MNRAS.332..915I} Ibata, R.~A., Lewis, G.~F., Irwin, M.~J., \& Quinn, T.\ 2002, \mnras, 332, 915 

  

\bibitem[Johnston et al.(2002)]{2002ApJ...570..656J} Johnston, K.~V., Spergel, D.~N., \& Haydn, C.\ 2002, \apj, 570, 656 

\bibitem[Just \& Pe{\~n}arrubia(2005)]{2005A&A...431..861J} Just, A., \& Pe{\~n}arrubia, J.\ 2005, \aap, 431, 861

\bibitem[Kamdar et al.(2019)]{2019arXiv190210719K} Kamdar, H., Conroy, C., Ting, Y.-S., et al.\ 2019, arXiv:1902.10719
  
\bibitem[Kandrup(1980)]{1980PhR....63....1K} Kandrup, H.~E.\ 1980, \physrep, 63, 1
  
\bibitem[Karl et al.(2015)]{2015MNRAS.452.2337K} Karl, S.~J., Aarseth, S.~J., Naab, T., Haehnelt, M.~G., \& Spurzem, R.\ 2015, \mnras, 452, 2337

\bibitem[\protect\citeauthoryear{Kashiyama \& Oguri}{2018}]{2018arXiv180107847K} Kashiyama K., Oguri M., 2018, arXiv, arXiv:1801.07847
  
\bibitem[\protect\citeauthoryear{Koushiappas \& Loeb}{2017}]{2017PhRvL.119d1102K} Koushiappas S.~M., Loeb A., 2017, PhRvL, 119, 041102
  


\bibitem[Lee(1968)]{1968ApJ...151..687L} Lee, E.~P.\ 1968, \apj, 151, 687 

\bibitem[Nelson \& Tremaine(1999)]{1999MNRAS.306....1N} Nelson, R.~W., \& Tremaine, S.\ 1999, \mnras, 306, 1

\bibitem[Ngan et al.(2016)]{2016ApJ...818..194N} Ngan, W., Carlberg, R.~G., Bozek, B., et al.\ 2016, \apj, 818, 194
  
\bibitem[Pe{\~n}arrubia et al.(2004)]{2004MNRAS.349..747P} Pe{\~n}arrubia, J., Just, A., \& Kroupa, P.\ 2004, \mnras, 349, 747
  
  
  
\bibitem[Pe{\~n}arrubia(2018)]{2018MNRAS.474.1482P} Pe{\~n}arrubia, J.\ 2018, \mnras, 474, 1482 ~(Paper I)

\bibitem[Pe{\~n}arrubia(2019)]{2019MNRAS.484.5409P} Pe{\~n}arrubia, J.\ 2019, \mnras, 484, 5409 ~(Paper II)

\bibitem[Prugniel \& Combes(1992)]{1992A&A...259...25P} Prugniel, P., \& Combes, F.\ 1992, \aap, 259, 25
  

\bibitem[Siegel et al.(2007)]{2007MNRAS.382..879S} Siegel, E.~R., Hertzberg, M.~P., \& Fry, J.~N.\ 2007, \mnras, 382, 879

  
  \bibitem[Smoluchowski(1916)]{1916ZPhy...17..557S} Smoluchowski, M.~V.\ 1916, Zeitschrift fur Physik, 17, 557 

  \bibitem[Spitzer(1987)]{1987degc.book.....S} Spitzer, L.\ 1987, Princeton, NJ, Princeton University Press, 1987, 191 p.,

    \bibitem[Springel et al.(2008)]{2008MNRAS.391.1685S} Springel, V., Wang, J., Vogelsberger, M., et al.\ 2008, \mnras, 391, 1685 

      
    \bibitem[van den Bosch et al.(1999)]{1999ApJ...515...50V} van den Bosch, F.~C., Lewis, G.~F., Lake, G., \& Stadel, J.\ 1999, \apj, 515, 50 
    \bibitem[van den Bosch(2017)]{2017MNRAS.468..885V} van den Bosch, F.~C.\ 2017, \mnras, 468, 885 


    \bibitem[van den Bosch \& Ogiya(2018)]{2018MNRAS.475.4066V} van den Bosch, F.~C., \& Ogiya, G.\ 2018, \mnras, 475, 4066
      
\bibitem[Wahde \& Donner(1996)]{1996A&A...312..431W} Wahde, M., \& Donner, K.~J.\ 1996, \aap, 312, 431 
  
  \bibitem[Weinberg(1989)]{1989MNRAS.239..549W} Weinberg, M.~D.\ 1989, \mnras, 239, 549 
  \bibitem[Weinberg(1993)]{1993ApJ...410..543W} Weinberg, M.~D.\ 1993, \apj, 410, 543
    
\bibitem[Yoon et al.(2011)]{2011ApJ...731...58Y} Yoon, J.~H., Johnston, K.~V., \& Hogg, D.~W.\ 2011, \apj, 731, 58 

\end{thebibliography}
\end{document}